\documentclass{optica-article}

\usepackage{graphicx}
\usepackage{float}
\usepackage{braket}
\usepackage{amsmath}
\usepackage{bm}

\journal{opticajournal}
\articletype{Research Article}
\usepackage{lineno}
\usepackage{datetime2}

\begin{document}

\title{Towards Quantum Telescopes: Demonstration of a Two-Photon Interferometer for Quantum-Assisted Astronomy}

\author{
Jesse Crawford \authormark{a},
Denis Dolzhenko \authormark{a},
Michael Keach \authormark{a},
Aaron Mueninghoff \authormark{b},
Raphael A. Abrahao \authormark{a}, 
Julian Martinez-Rincon \authormark{a},
Paul Stankus \authormark{a},
Stephen Vintskevich\authormark{c},
Andrei Nomerotski \authormark{a}
}

\address{\authormark{a}Brookhaven National Laboratory, Upton NY 11973, USA\\
\authormark{b}Stony Brook University, Stony Brook, NY 11794, USA\\
\authormark{c}Independent researcher, Ras Al Khaimah, United Arab Emirates}

\email{\authormark{*}e-mails for correspondence: anomerotski@bnl.gov}

\begin{abstract*}


Classical optical interferometery requires maintaining live, phase-stable links between telescope stations.  This requirement greatly adds to the cost of extending to long baseline separations, and limits on baselines will in turn limit the achievable angular resolution.  Here we describe a novel type of two-photon interferometer for astrometry, which uses photons from two separate sky sources and does not require an optical link between stations.
Such techniques may make large increases in interferometric baselines practical, even by orders of magnitude, with corresponding improvement in astrometric precision benefiting numerous fields in astrophysics.  We tested a benchtop analogue version of the two-source interferometer and unambiguously observe correlated behavior in detections of photon pairs from two thermal light sources, in agreement with theoretical predictions.  This work opens new possibilities in future astronomical measurements.

\end{abstract*}

\section{Introduction and Basic Concepts}


Classical optical Michelson interferometers collect photons from a sky source into two or more
sub-apertures, and these are then transported through optical links and brought into interference at a common point. For a pair of sub-apertures separated by a baseline $B$ the measured interference pattern is sensitive to features of the source with angular size 
 $\delta\theta\sim\lambda/B$, where $\lambda$ is the photon wavelength. The optical link between the stations must meet demanding requirements, with lengths remaining stable to within a fraction of a wavelength. This makes the Michelson interferometry expensive and difficult to extend to long baselines, and this in turn limits the achievable resolution \cite{Pedretti2009, tenBrummelaar2005}. 

The use of quantum optics to improve the precision of astronomical measurements is a long-desired goal of both the optical and astronomical communities. In particular, the seminal Gottesman-Jennewein-Croke (GJC) proposal \cite{Gottesman2012} attracted attention as a way to build a quantum-enhanced telescope, i.e. a very-long-baseline interferometer enabled by the use of quantum optical effects. However, the GJC proposal is dependent on the use of quantum repeaters \cite{RMP2011_quantum_repeaters,IEEErevQRepeaters,carlson202repeater,SPIE_Gottesman_QTelescope}, a technology which still requires a substantial amount of development,  limiting  practical implementations of the technique. 
Alternatively, the proposed Stankus-Nomerotski-Slosar-Vintskevich (SNSV) approach \cite{Stankus2022} relies on enhancements of the Hanbury Brown and Twiss (HBT) effect.
It simplifies requirements for an optical quantum-assisted astronomical measurement, providing a practical pathway to achieve the goal of more precise astrometric measurements for astronomical objects, and is based on the quantum interference phenomena, while avoiding employment of quantum repeaters. Other theoretical techniques for improved resolution on imaging the starlight were also introduced and some proof-of-principle demonstrations were successfully performed \cite{TsangPRL2011,Tsang_review2019,Pearce2017optimalquantum,UQ2019,Singapore_OE2016}.

\begin{figure}[!htp]
\begin{center}
\includegraphics[width=0.99\linewidth]{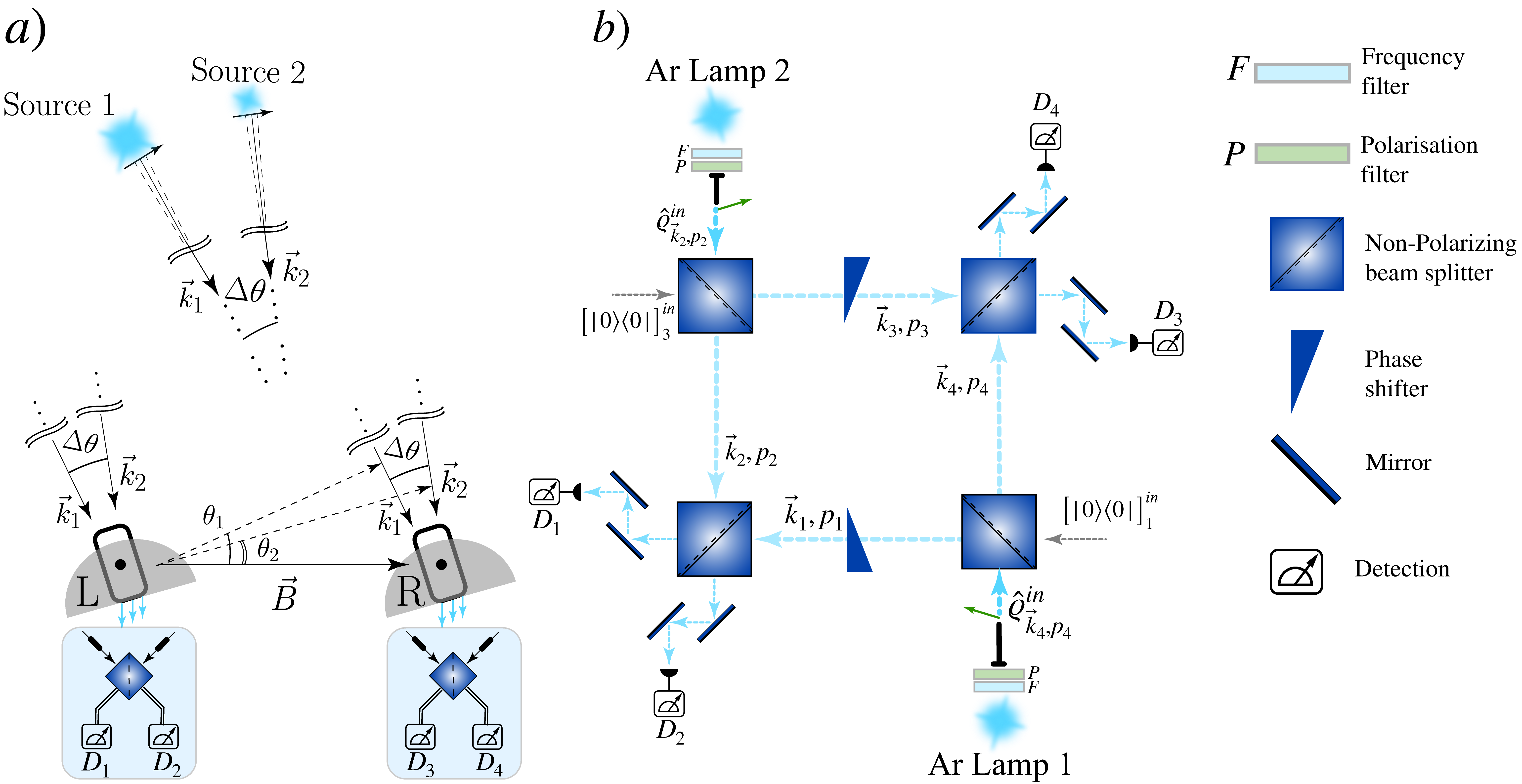}
\caption{$(a)$ Concept of the proposed Stankus-Nomerotski-Slosar-Vintskevich (SNSV) two-photon interferometer, which interferes and detects photons, shown as flat waves, from two astronomical sources. $(b)$ Equivalent scheme for the tabletop implementation. The argon lamps 1 and 2 indicate the two ports associated with input spatial and polarization modes. 
Each output detection port $D_{1}, D_{2}, D_{3},$ and $D_{4}$ corresponds to detectors in two observation stations $L$ and $R$ in (a).
}
\label{fig:idea}
\end{center}
\end{figure}

Here we describe a proof-of-principle experiment based on the SNSV scheme 
\cite{Stankus2022, Nomerotski20, Keach2022}, shown schematically in the left part of Figure \ref{fig:idea}. 
In brief, photons are collected from two sky sources in two observing stations. Light from the sources is directed into beamsplitters for interference. Correlations of photon detections between the four beamsplitter outputs are sensitive to the relative phase difference of incoming photons from the two sources. From this the opening angle between the sources can be determined, allowing for longer baselines when compared to the classical interferometry, since it entirely removes the need for a physical optical path between the stations. Therefore, such an approach could improve astrometric precision. 

The measured coincidence rates of the outputs can be described by a second-order intensity correlation function $\Gamma_{jk}$ which has the following simplified form:
\begin{equation}\label{eq:prob_eq}
    \Gamma_{jk} = A + B\cos(\delta_{j}-\delta_{k})
\end{equation}
where A and B are the coefficients dependent on the photon polarization and coherence properties, and $\delta_{j,k}$ are the phases for given modes.

We provide a detailed theoretical derivation of the intensity correlation function in quasi-monochromatic approximation in the Supplement focusing on the polarization degrees of freedom, and on the relationship between the mode indistinguishability and interference effects. There, we predict that the output channels from Figure~\ref{fig:idea} should form pairs such that channels 1\&3 should be anti-correlated with channel pairs 1\&4 and 2\&3 but correlated with 2\&4, while the channel pair 1\&4 would be correlated with the channel pair 2\&3 but anti-correlated with 1\&3 and 2\&4. We also predict that coincidences of the channel pairs 1\&2  and 3\&4  would have a  stationary rate without oscillations, a non-trivial correlation pattern.

The basis of the proposed technique relies on two phenomena: the Hanbury Brown-Twiss (HBT) \cite{HBT_original} and Hong-Ou-Mandel (HOM) \cite{HOM_effect} effects. We expect the two photons to bunch within their coherence time as in the HBT effect, resulting in the HBT peaks in the time difference distributions. In addition to this, we predict that the population of the HBT peak would depend on the relative phase difference between the two photons. The HOM effect would play a role when correlations between two outputs of the same beamsplitter are considered, leading to two indistinguishable photons to coalesce at a beamsplitter output. Note that the presented scheme has similarities with the Noh-Fougères-Mandel experiments \cite{NohMandel1991,NohMandel1992a,NohMandel1992b,NohMandel1993} but, in contrast, it is focused on the second-order correlation analysis, which describes coincidences between pairs of detectors.
The optical effects discussed above is a manifestation of the interference of two indistinguishable photons. To be indistinguishable the photons must have similar frequencies and arrive at the beamsplitters at the same time. These arguments establish requirements on the timing and spectral resolution for the interferometer instrumentation \cite{Nomerotski20, Sensors2020_Nomerotski, PRA_Jordan2022, Bouchard_HOM_review}. 

The SNSV technique has parallels with the GJC approach \cite{Gottesman2012}, which employs a path-entangled source of single photons distributed between two observing stations, and its variations \cite{harvard1, harvard2, Brown2021, Brown2022,FIO2021Kwiat,OSAOptica2_Kwiat_2022}. In the SNSV proposal, photons provided by the second thermal light source would replace photons from a single photon source, typically a spontaneous parametric down conversion (SPDC) source \cite{SPDC_general}. While an on-sky light source is inherently thermal, we note that this has important practical advantages, compared to the above-mentioned quantum source, such as the complete absence of the direct optical link between the stations and uniformity of instrumentation for observations.

The two-photon interferometer presented here allows for precision improvement, in principle, by orders of magnitude, which could benefit numerous fields in cosmology and astrophysics. There are many scientific opportunities that would benefit from substantial improvements in astrometric precision such as testing theories of  gravity  by  direct  imaging  of black  hole  accretion  discs, precision parallax for the cosmic distance ladder, mapping microlensing events, peculiar motions and dark matter; see Ref. \cite{Stankus2022} for a more comprehensive discussion. As a numerical example, it was determined in Ref. \cite{Stankus2022} that a nominal precision on the order of 10~$\mu$as on the opening angle between two bright stars in a single night's observation could be reached.

In the following, we start with discussing the benchtop experimental setup of the interferometer in Section \ref{sec:setup}. We then focus on the methods and results in Section \ref{sec:methods}. 
Section \ref{sec:discussion} provides a discussion of the results. Lastly, we present the conclusions and the future outlook in Section \ref{sec:conclusions}. Detailed theoretical considerations and extended experimental setup description are presented in the Supplement.

\section{Experimental Setup}
\label{sec:setup}

The experimental setup utilized in the measurements is shown in the right part of Figure \ref{fig:idea}. Two argon lamps with isolated 794.82~nm spectral lines were used as sources of thermal photons. The photons were directed to four 50:50 non-polarizing beamsplitters to arrange interference as in the original scheme in the left part of Figure \ref{fig:idea}. The photon phases remained stable to the environmental disturbances for extended periods of time and could be adjusted deterministically with specialized phase shifters to study the phase dependence. The shifters were implemented as small angle glass wedges, which can be moved laterally in fine steps. The four outputs of the interferometer were instrumented with fast single-photon detectors. Two types of detectors were used, the single photon avalanche detectors (SPAD) \cite{Gasparini17, Perenzoni16, Lee18} and superconducting nanowire single-photon detectors (SNSPD) \cite{Divochiy08, Zhu20, Korzh20}, both with temporal resolution of the order of 100~ps. The SPAD/SNSPD digital output signals were then read out by a TDC module. A detailed description of the setup is provided in the Supplement.

We employed the following experimental procedure. Firstly, one of the phase shifters was moved in small steps by a distance of about 0.45 mm over the duration of 15 minutes, which corresponded to a shift by five wavelengths, followed by a pause of two minutes. Then, the second phase shifter was moved in the same manner followed by another two-minute pause. The total duration of the undisturbed measurement was about 35 minutes. The photon time-stamps of four output channels were continuously logged on disk for post-processing. 
A variety of runs were performed with two different detector types and varying polarizer configurations. These configurations included experiments with unpolarized photons, when the polarizers were removed from the beam paths, and experiments with different respective polarizations for the two lamps.  

\section{Methods and Results}
\label{sec:methods}

The main goal of the analysis was to determine the dependence of two-photon correlations on the relative phase of the photons. Algorithms were developed to condition the raw data by removing the afterpulses, then to find coincidences of photon pairs in different channels to identify the HBT peaks, and to determine the dependence of the peak population on the photon phase. 

\subsection{HBT Peaks}

The HBT peaks appear in the time difference distributions of channel pairs due to the two-photon interference yielding the photon bunching \cite{HBT_original}. The characteristic shape of the distribution is determined by the convolution of the corresponding photon coherence time due to the spectral width of the argon line, and timing resolution \cite{Nomerotski20}. We study this effect by analyzing the distribution of photon detection time differences by combining various detector channels. The analysis algorithm searches for pairs of single photons detected within a $\pm$ 20 ns window from each other. 

In previous work \cite{Nomerotski20}, we also determined that the timing resolution is the predominant contribution to the HBT peak temporal width, so we model the HBT peaks with a normalized $g^{(2)}$ autocorrelation function of this form:
\begin{equation}
    g_{ij}^{(2)}(\Delta t, t_{0},p) = 1 +  V_{HBT}e^{-\frac{\left(\Delta t - t_{0}\right)^2}{2\sigma^2}}, \ p \in \{VV, VH\}, \ ij\in\{12,13,14,23,24,34\},
\label{eq:gauss}
\end{equation}
where $V_{HBT}$ is the visibility of HBT effect, $\Delta t$ is the time difference with an offset $t_0$, $\sigma$ is the standard deviation, $p$ is the polarization, and indices $i$ and $j$ are labeling the detector pairs; see Eqs. (S10) - (S14) in the Supplement for the detailed derivation.

\begin{figure}[htp!]
    \centering
    \includegraphics[width=.45\linewidth]{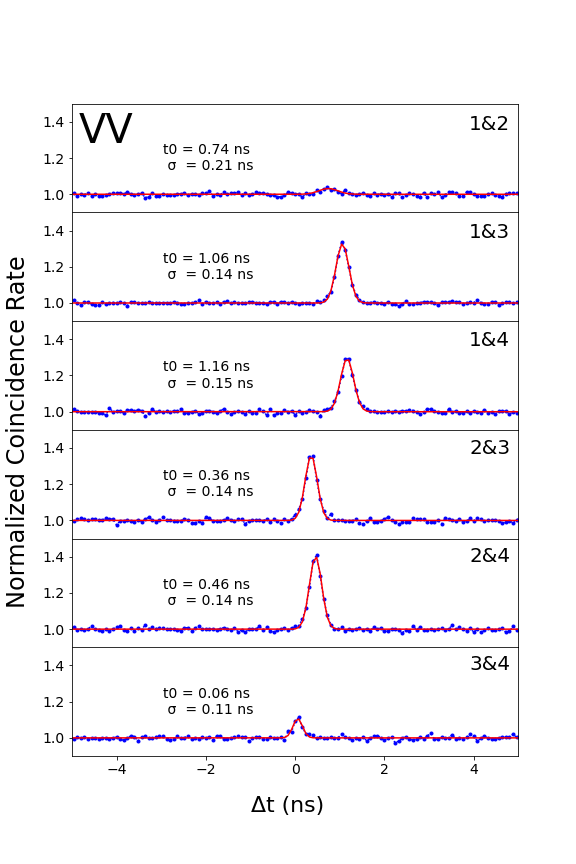}
    \includegraphics[width=.45\linewidth]{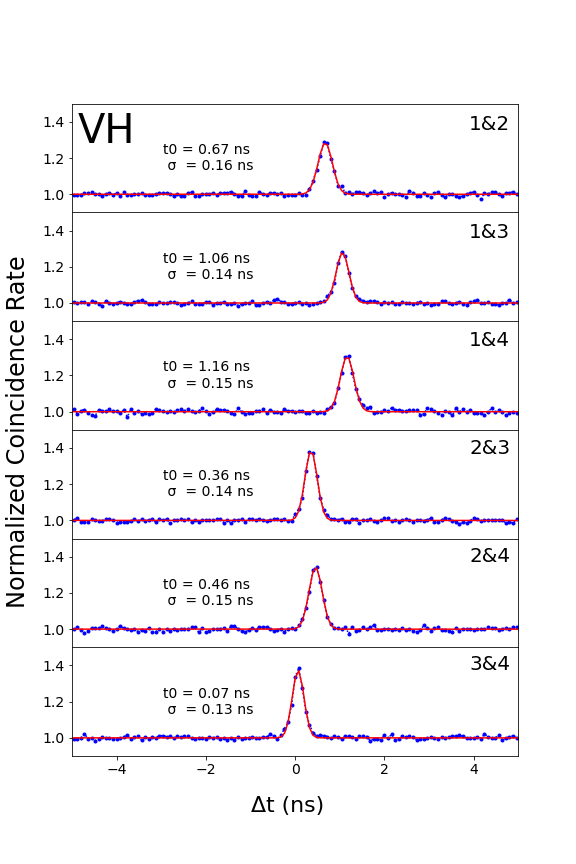}
    \vspace{-0.5cm}
    \caption{Normalized coincidence count rates of two-photon detections as a function of the time difference between them, $\Delta$t, for different output channel combinations and different input polarizations, as labeled. The peak in each case indicates the enhanced correlation between two photons, calibrating where simultaneous pairs will appear. 
    Left: results with both polarizers aligned vertically, called VV configuration. Right: results with one polarizer aligned vertically and the other polarizer aligned horizontally, called VH configuration. The parameters $t_0$ and $\sigma$ are the time offset and the Gaussian fit width, respectively, according to Equation~\ref{eq:gauss}. The peaks are signatures of the Hanbury Brown-Twiss (HBT) effect, appearing due to the photon bunching.}
    \label{fig:HBTpeaks}
\end{figure}

As expected, prominent peaks are seen for most of the six combinations of channels, as shown in Figure \ref{fig:HBTpeaks}. 
The left and right parts of Figure \ref{fig:HBTpeaks} show the HBT peaks with vertical-vertical, or VV, and vertical-horizontal, or VH, as the orientation of the polarizers at the two inputs. Time offsets in the peak positions are due to the varying paths in different channels including small differences in the length of optical fibers. Note that for the VH case the interference is happening for two photons from the same source and the VH photon pairs from the different sources do not contribute. We discuss properties of the HBT peak distributions in Section \ref{sec:discussion}.

\subsection{Coincidence rates}
\label{sec:coinc.rates}
\begin{figure}[htp]
\begin{tabular}{c} 
\includegraphics[width=0.45\linewidth]{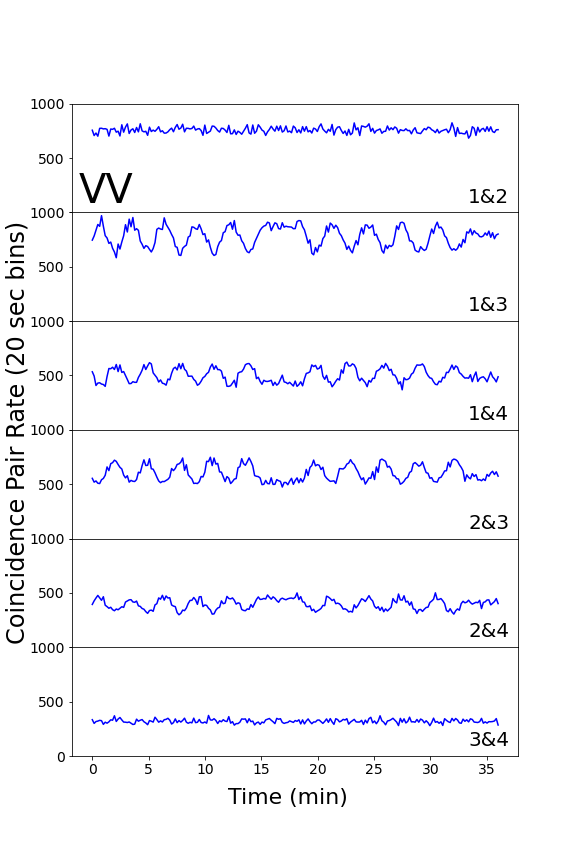}
\includegraphics[width=0.45\linewidth]{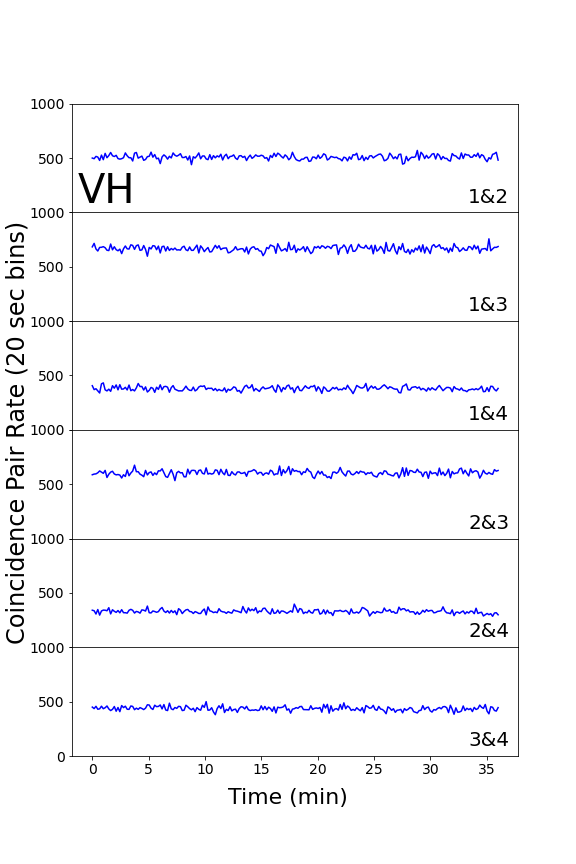}
\end{tabular}
\vspace{-0.5cm}
\caption[PairRates] 
{ \label{fig:PairRates} 
        Number of two-photon detections within a $\pm$ 1.5$\sigma$ window about $t_0$ of the HBT peak in 20 sec time bins plotted versus time for the SNSPD data set with polarizers. The time accounts for the real time during which the phase shifters were slowly moving a preset distance corresponding to five wavelengths, then  paused for two minutes, then moved again for five wavelengths. The graphs show results for six combinations of channel pairing, 1\&2, 1\&3, 1\&4, 2\&3, 2\&4 and 3\&4. Left: Results with VV configuration of polarizers. Right: Results with VH configuration of polarizers. See the text for discussion.}
\label{fig:OSCL}
\end{figure}

As the first step, the HBT peaks were fit with a Gaussian function to determine the peak width $\sigma$ and the central value of the entire 35~min dataset. Then the coincidence rate for channel pairs was determined using two techniques. In the first, simple approach a window of $\pm 1.5 \sigma$ around the HBT peak central value was selected, where $\sigma$ was taken from the corresponding fit of the peak. Then the number of entries in this window in a predefined time bin, typically 10~-~30 sec, was determined and plotted as a function of time as shown in Figure \ref{fig:OSCL} for all six channel combinations for the SNSPD data set with polarizers. 
The left and right parts of Figure \ref{fig:OSCL} show, respectively, the number of coincidences in the 20 sec time bins with  VV and VH polarizer orientation at the interferometer inputs. 
It is expected that these time trending plots will have oscillatory behavior in the VV case due to the advance of the photon phase caused by the phase shifters as predicted in Equation \ref{eq:prob_eq}. We indeed observe this behavior but defer a detailed discussion of the main features of these measurements to Section \ref{sec:discussion}.

\begin{figure} [htp]
    \centering
    \includegraphics[width=0.55\linewidth]{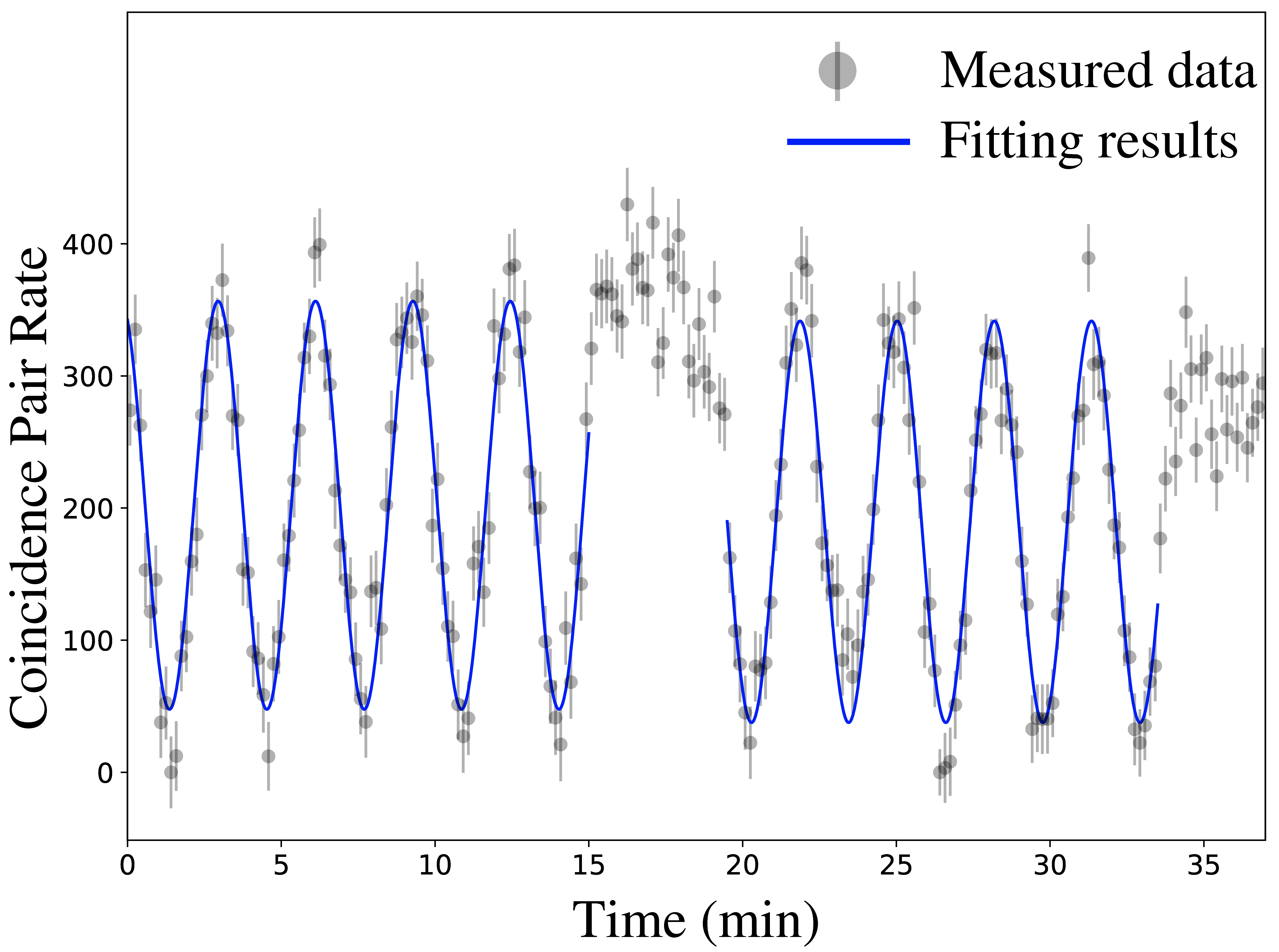}
    \caption{Two-photon coincidences count rates for the oscillations for channels 1\&3 in the SNSPD data set with with VV polarization fit to Equation \ref{eq:cos}. The coincidences rates were determined fitting a Gaussian peak in each 10 sec time bin. Data points are presented together with one standard deviation error bars.}
    \label{fig:cosfit}
\end{figure}

In a more sophisticated approach, the HBT peak in each predefined, sequential time bin was fit with a Gaussian profile, only allowing to vary the peak amplitude and background detection level. The peak center $t_0$ and standard deviation $\sigma$ of these fits were held constant at the values determined from the initial fit of the overall HBT peak. Then the area under each of these Gauss peaks within $\pm 1.5 \sigma$ of the peak center was determined, effectively subtracting the background detection rate, and plotted as a function of time. We show an example of such fit in Figure \ref{fig:cosfit}.
Mathematically this result can be described as the convolution of the second-order correlation function 
with a filtering function, see Eqs.(S13) and (S14) in Supplement. All fits in this method were made using LMFIT \cite{lmfit}, which was also used to explicitly calculate uncertainties on the parameter best-fit values. This technique should give better statistical accuracy since it uses more statistics for the flat background and can also account for a slow drift of the flat background. 

To determine the visibility and relative phase, the oscillatory behavior of trending plots due to the phase evolution was fit with a cosine function: 
\begin{equation}
\Gamma_{ip_{i}jp_{j}}(t) = \langle A \rangle_{ip_{i}jp_{j}} + \langle B\rangle_{ip_{i}jp_{j}}\cos{\left(\frac{2\pi}{T}t-\Delta\delta_{ij} \right)}.
\label{eq:cos}
\end{equation}
where $\langle B\rangle$ denotes the signal amplitude, $T$ is the characteristic period of slow phase adjustments by the phase shifters, $t$ is the data time stamp, $\Delta\delta_{ij} \equiv \delta_{i} - \delta_{j}$ is the constant relative phase, and $\langle A\rangle$ is the background level. Labels $p_{i} \in \{H,V\}$ and $p_{j} \in \{H,V\}$ are labels of polarization modes tailored to spatial modes $i$ and $j$ respectively. We define the corresponding visibility as: 
\begin{equation}
    {\mathcal{V}_{OSC}}=\frac{\langle B\rangle_{ip_{i}jp_{j}}}{\langle A \rangle_{ip_{i}jp_{j}}}.
    \label{vis_osc}
\end{equation}

\section{Analysis and Discussion}
\label{sec:discussion}

Oscillations in the trending plots were fit using a cosine function as explained above, from which the main parameters were extracted. These parameters included the oscillation period, visibility and phase. Out of these, the latter two, visibility and relative phase, are critical parameters, unambiguously predicted by the theory.

\subsection{Oscillations in Coincidence Rate}

We show the coincidence rates for all channel combinations in Figure \ref{fig:OSCL}. As previously shown in Equation \ref{eq:prob_eq}, the channel pairs on the opposite interferometer arms (pairs 1\&3, 1\&4, 2\&3, and 2\&4) are expected to oscillate in or out of phase, while channel pairs on the same interferometer arm (pairs 1\&2, 3\&4) are not expected to oscillate.
Since for the HOM effect only indistinguishable photons are supposed to interfere, we do not expect any oscillations in the VH configuration of polarizations.

\begin{figure} [htp!]
    \centering
    \includegraphics[width=0.85\linewidth]{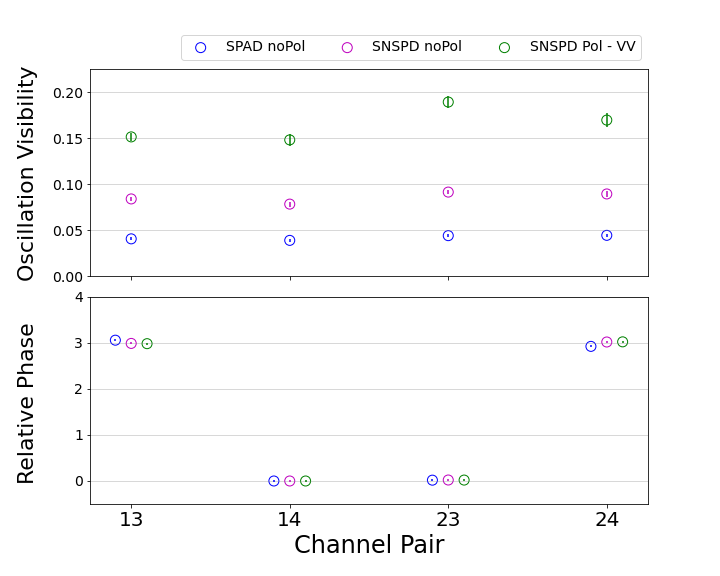}
    \caption{Top: Visibility of oscillations for the two-photon coincidences count rates. Bottom: Relative phases, in radians, calculated with respect to the channel pair 1\&4 phase. Colors correspond to different datasets. Pol and noPol are the VV poparized and unpolarized datasets, respectively. Data points are presented together with one standard deviation error bars.}
    \label{fig:cos_vis}
\end{figure}

Only the first five oscillation periods were used for the cosine fits in these results
extracting the relative phase, which was calculated with respect to the channel pair 1\&4 phase, and the visibility, calculated using Eq.~\ref{vis_osc}. This analysis was performed for several datasets: VV polarized and unpolarized datasets with SNSPD detectors, and unpolarized dataset with SPAD detectors. We summarize the measured visibility for the cross-station channel combinations and for the datasets in the top part of Figure \ref{fig:cos_vis} and the relative phases, in radians, in the bottom part of Figure \ref{fig:cos_vis}. The measured parameters all behave as expected. As can be seen the visibility was higher for the polarized dataset, as expected, by approximately a factor of two compared to the unpolarized dataset. In the unpolarized case, the visibility of oscillations for the SPAD detectors were consistently smaller compared to the visibility of the peaks for the SNSPD detectors. This can be explained by the inferior timing resolution of SPAD detectors.

\begin{figure} [htp!]
\centering
\includegraphics[width=0.75\linewidth]{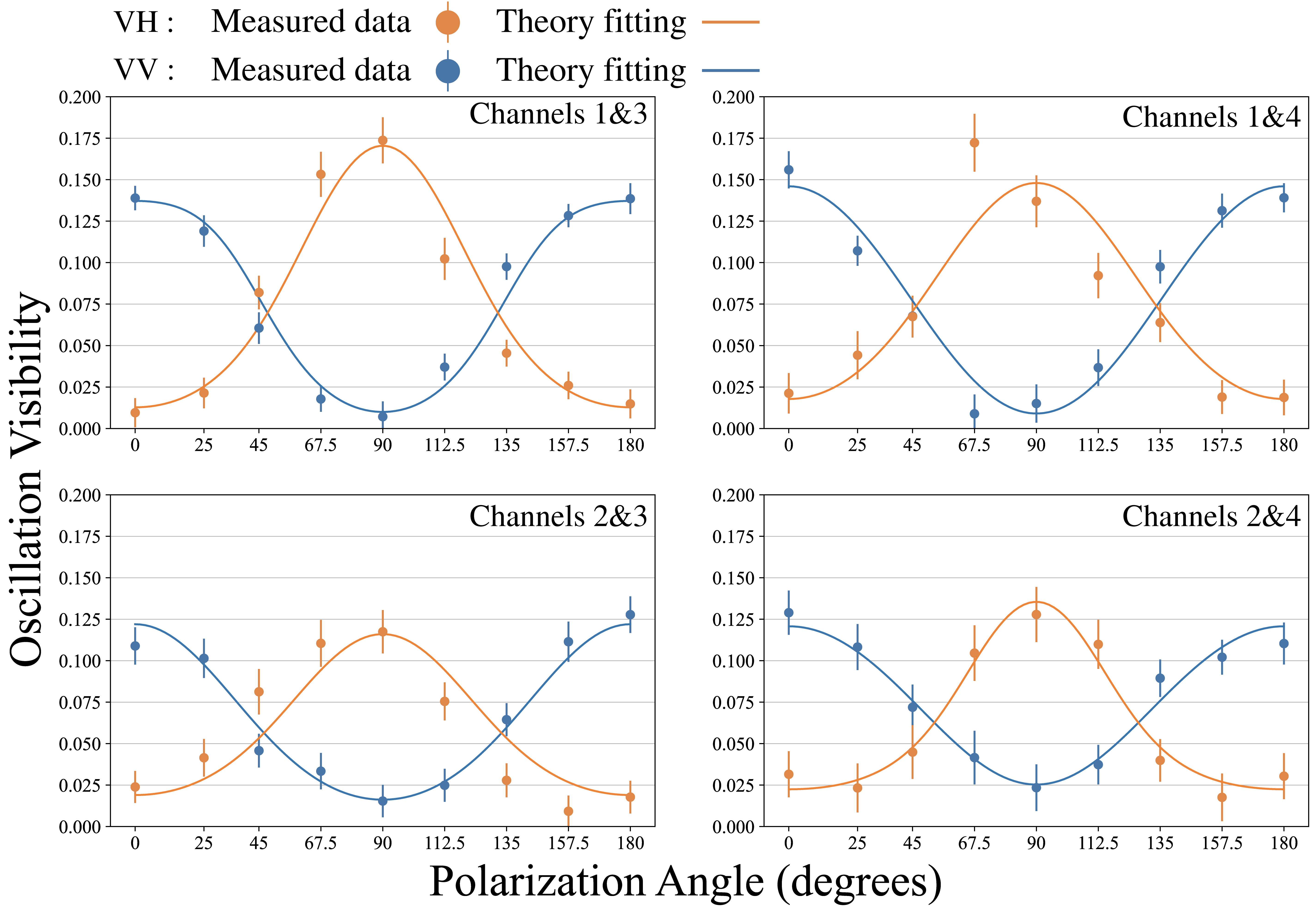}
    \caption{Visibility of two-photon detections as a function of polarization angle, for channel pairs 1\&3, 1\&4, 2\&3, and 2\&4. Here, one input polarization angle was varied in 22.5$^\circ$ increments from 0$^\circ$ to 180$^\circ$ starting from either VV or VH configuration, while the other polarizer was not moved. Different colors refer to different configurations as indicated. Data points are presented together with one standard deviation error bars. The measured experimental data is fitted by Eq.~\ref{eq:vis_theory_theta_1}. 
    To fit the results we added to the fit function a small constant offset,
    which described the accidental coincidences counts.
    }
    \label{fig:Visibility}
\end{figure}

We also checked the behavior of the interferometer for varying polarizations of the two beams. Figure \ref{fig:Visibility} shows the visibilities of two-photon detections as a function of polarization angle, for channel pairs 1\&3, 1\&4, 2\&3, and 2\&4. Here, one input polarization angle was varied in 22.5$^\circ$ increments from 0$^\circ$ to 180$^\circ$ starting from either VV or VH configuration, while the other polarizer was not moved. The dependence on the relative angle for the two polarizations was changing in anti-phase for these two cases, as expected. We compare the experimental data to the theoretical predictions, as in Eq.~\ref{eq:vis_theory_theta_1}, derived in Supplement,
and find a good agreement.
\begin{eqnarray}
    {V_{VV}(\theta)} = \frac{r_{VV} \sin^{2}(\theta)}{1 + r_{VV}\sin^{2}{\theta}+r_{VV}^{2}\sin^{4}{\theta}+r_{HV}^{2}\cos^{4}{\theta} + r_{HV}\cos^{2}{\theta} + r_{HV}r_{VV}\sin^{2}{\theta}\cos^{2}{\theta}} \nonumber \\ 
    {V_{VH}(\theta)} = \frac{r_{HH} \cos^{2}(\theta)}{1 + r_{HH}\cos^{2}{\theta}+r_{HH}^{2}\cos^{4}{\theta}+r_{VH}^{2}\sin^{4}{\theta} + r_{VH}\sin^{2}{\theta} + r_{HH}r_{VH}\sin^{2}{\theta}\cos^{2}{\theta}}, 
    \label{eq:vis_theory_theta_1}
\end{eqnarray}
where $r_{ij}$ are the normalized rates defined in Supplement.

\subsection{ HBT peak visibility, cancellation of HOM and HBT effects}
\label{sec:cancellation.polarization}

\begin{figure} [ht]
    \centering
    \includegraphics[width=0.6\linewidth]{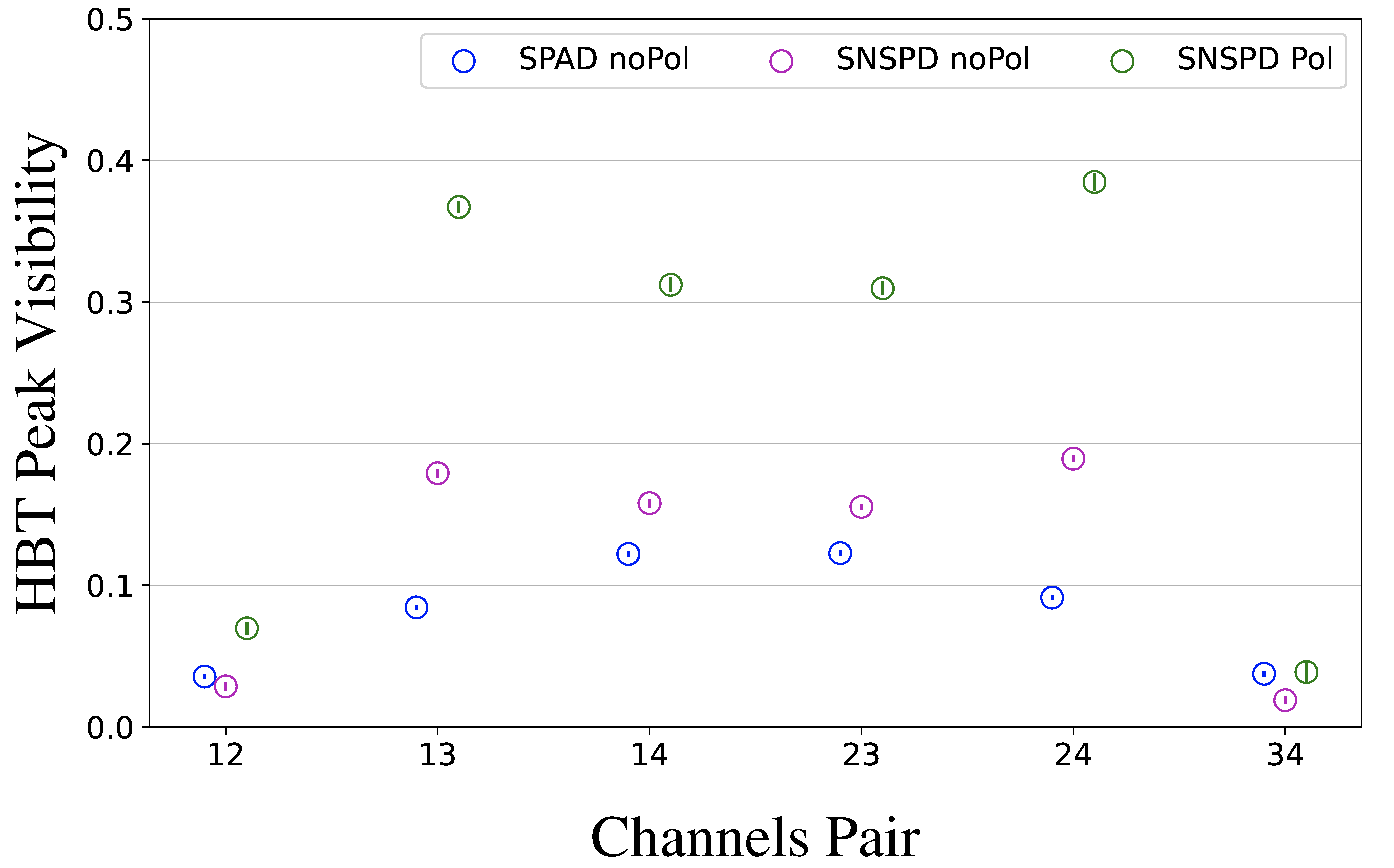}
\caption{
Visibility of the HBT peaks presented in Figure \ref{fig:HBTpeaks} for various datasets. Colors correspond to different datasets. Pol and noPol are the VV poparized and unpolarized datasets, respectively. Data points are presented together with one standard deviation error bars.}
    \label{fig:HBTvis}
\end{figure}

As already discussed in Section \ref{sec:methods} the HBT peaks were fit with a Gaussian function. The resulting timing resolution ($\sigma$) was found to be equal to  280~ps and 140~ps, respectively for SPAD and SNSPD detectors. This is in agreement with expected timing resolution of those detectors assuming a time difference measurement for two independent photons. 
The HBT peak visibility was introduced in Eq.~\ref{eq:gauss}. A  graph summarizing the visibilities for all combinations of channel pairs in the main datasets is shown in Figure \ref{fig:HBTvis}.  Similarly to the oscillation visibility, the HBT visibility was significantly higher for the polarized case, decreasing for the unpolarized case and for the SPAD detectors, which have worse timing resolution.

We also note that the HBT peak visibility is almost zero for the 1\&2 and 3\&4 pair combinations in the VV configuration. These particular channels evaluate coincidences of two photons exiting the two opposite sides of the same beamsplitter so their coincidence rate will have a dip due to the Hong-Ou-Mandel (HOM) effect. This will cancel out the photon bunching peak due to the HBT effect. However, the results in Figures \ref{fig:HBTpeaks} and \ref{fig:HBTvis} show it's not an exact cancellation, which is likely due to the non-ideal equalization of the interferometer arms, also a valid explanation for the spread of visibility for the middle four channel combinations in Figures \ref{fig:cos_vis} and \ref{fig:HBTvis}.

\section{Conclusions and Outlook}
\label{sec:conclusions}
In this work we implemented a proof-of-principle demonstration of the SNSV proposal for quantum-assisted astronomy \cite{Stankus2022} and described bench-top experiments with a two-photon interferometer.
The results, in particular, the observed phase dependence, confirm the predicted functionality of the proposed instrument, suggesting that it is a viable experimental approach that can improve the astrometric precision. 

The next steps in the exploration of the approach would be observations of on-sky light sources. We note that the phase shifting performed in the above experiments and corresponding oscillations in the coincidence rates are directly analogous to the Earth rotation fringe shifting \cite{Zhi2022}. Due to Earth's rotation, the effective baseline between the two stations change, which induces changes in the interference patterns, in particular, in the frequency of corresponding fringe oscillations. The frequency value is predicted in \cite{Stankus2022} to be proportional to the opening angle between the stars. The uncertainty of the fringe rate has more favorable scaling with observation period than simple photon statistic methods, so it can be considered a promising observable for the quantum astrometry approach.

The star spectra are typically broadband, so another obvious extension of the technique is spectroscopic binning, which would allow us to estimate the observables in numerous spectral bins. Each bin would act as an independent experiment, so the sensitivity of the interferometer would improve as $\sqrt{N}$, where $N$ is the number of spectral bins. The sensitivity also improves with a larger number of stations and better timing resolution, as discussed in Ref.~\cite{Stankus2022}.

Though we operate here with the thermal states of light, which have the classical Gaussian quasiprobability distribution, we consider a quantum description of these phenomena, as in this work, to be very instructive. It emphasizes the role of mode (or path) indistinguishability in the quantum interference phenomena. Moreover, this description can be extended further by employing the Quantum Continuous Variable (CV) formalism \cite{Serafini2017}. The CV formalism will play an important role in expanding the presented technique to multiple observing stations assisted by auxiliary ground-based quantum states such as squeezed states and non-trivial quantum channels 
\cite{Xia21,VintskevichPRA2019,Brady2022,wu2022entanglement,Zhang2019,cox2022transceiver}, providing novel opportunities in astronomy for extraction of valuable information. At the same time, the complexity of analytical descriptions and data processing would grow tremendously with an increased number of stations and auxiliary quantum states involved in the measurements, including entangled ones \cite{Gottesman2012,harvard1,harvard2}, so  an important future research goal is to provide a theoretical description of such expansions based on both entities, the Gaussian states and quantum channels, powered by machine-learning methods \cite{Krenn_review, vintskevich2022}.

In summary, we built and characterized a tabletop prototype of a two-photon interferometer, which could improve the astrometric precision by orders of magnitude, by means of enabling extra long baselines between observing stations. The implemented interferometer allowed us to test the important features of the SNSV proposal, in particular to demonstrate that the relative phase of two photons from two independent thermal sources has a direct effect on their bunching due to the HBT effect. The approach demonstrated here is technically feasible with existing technologies of single photon detection \cite{Nomerotski20, Zhi2022} and allows us to move towards measurements with on-sky sources. This work represents a major step towards quantum-assisted astronomy.

\section{Acknowledgments}
This work was supported by the U.S. Department of Energy QuantISED award and BNL LDRD grants 19-30 and 22-22. 
A.M. acknowledges support under the Lourie Fellowship from the Stony Brook University Department of Physics and Astronomy. We are grateful to Jonathan Schiff and Rom Simovitch for the software development and support.

\section{Disclosures}
The authors declare no conflicts of interest.


\bibliography{references} 

\begin{thebibliography}{10}
\newcommand{\enquote}[1]{``#1''}

\bibitem{Pedretti2009}
E.~Pedretti, J.~D. Monnier, T.~ten Brummelaar, and N.~D. Thureau,
  \enquote{Imaging with the {CHARA} interferometer,} {\protect\JournalTitle{New
  Astronomy Reviews}} \textbf{53}, 353--362 (2009).

\bibitem{tenBrummelaar2005}
T.~A. ten Brummelaar, H.~A. McAlister, S.~T. Ridgway, J.~W.~G.~Bagnuolo, N.~H.
  Turner, L.~Sturmann, J.~Sturmann, D.~H. Berger, C.~E. Ogden, R.~Cadman
  \emph{et~al.}, \enquote{First results from the {CHARA} array. {II}. a
  description of the instrument,} {\protect\JournalTitle{The Astrophysical
  Journal}} \textbf{628}, 453--465 (2005).

\bibitem{Gottesman2012}
D.~Gottesman, T.~Jennewein, and S.~Croke, \enquote{Longer-baseline telescopes
  using quantum repeaters,} {\protect\JournalTitle{Phys. Rev. Lett.}}
  \textbf{109}, 070503 (2012).

\bibitem{RMP2011_quantum_repeaters}
N.~Sangouard, C.~Simon, H.~de~Riedmatten, and N.~Gisin, \enquote{Quantum
  repeaters based on atomic ensembles and linear optics,}
  {\protect\JournalTitle{Rev. Mod. Phys.}} \textbf{83}, 33--80 (2011).

\bibitem{IEEErevQRepeaters}
W.~J. Munro, K.~Azuma, K.~Tamaki, and K.~Nemoto, \enquote{Inside quantum
  repeaters,} {\protect\JournalTitle{IEEE Journal of Selected Topics in Quantum
  Electronics}} \textbf{21}, 78--90 (2015).

\bibitem{carlson202repeater}
E.~K. Carlson, \enquote{The key device needed for a quantum internet,}
  {\protect\JournalTitle{Physics}} \textbf{13}, 104 (2020).

\bibitem{SPIE_Gottesman_QTelescope}
D.~Gottesman, \enquote{{Quantum telescopes},} in \emph{Optical and Infrared
  Interferometry and Imaging VII,}  vol. 11446 P.~G. Tuthill, A.~M{\'e}rand,
  and S.~Sallum, eds., International Society for Optics and Photonics (SPIE,
  2020), p. 1144615.

\bibitem{Stankus2022}
P.~Stankus, A.~Nomerotski, A.~Slosar, and S.~Vintskevich, \enquote{Two-photon
  amplitude interferometry for precision astrometry,}
  {\protect\JournalTitle{The Open Journal of Astrophysics}} \textbf{5} (2022).

\bibitem{TsangPRL2011}
M.~Tsang, \enquote{Quantum nonlocality in weak-thermal-light interferometry,}
  {\protect\JournalTitle{Phys. Rev. Lett.}} \textbf{107}, 270402 (2011).

\bibitem{Tsang_review2019}
M.~Tsang, \enquote{Resolving starlight: a quantum perspective,}
  {\protect\JournalTitle{Contemporary Physics}} \textbf{60}, 279--298 (2019).

\bibitem{Pearce2017optimalquantum}
M.~E. Pearce, E.~T. Campbell, and P.~Kok, \enquote{Optimal quantum metrology of
  distant black bodies,} {\protect\JournalTitle{{Quantum}}} \textbf{1}, 21
  (2017).

\bibitem{UQ2019}
L.~A. Howard, G.~G. Gillett, M.~E. Pearce, R.~A. Abrahao, T.~J. Weinhold,
  P.~Kok, and A.~G. White, \enquote{Optimal imaging of remote bodies using
  quantum detectors,} {\protect\JournalTitle{Phys. Rev. Lett.}} \textbf{123},
  143604 (2019).

\bibitem{Singapore_OE2016}
Z.~S. Tang, K.~Durak, and A.~Ling, \enquote{Fault-tolerant and finite-error
  localization for point emitters within the diffraction limit,}
  {\protect\JournalTitle{Opt. Express}} \textbf{24}, 22004--22012 (2016).

\bibitem{Nomerotski20}
A.~Nomerotski, P.~Stankus, A.~Slo{\v{z}}ar, S.~Vintskevich, S.~Andrewski, G.~A.
  Carini, D.~Dolzhenko, D.~England, E.~V. Figueroa, S.~Gera, J.~Haupt,
  S.~Herrmann, D.~Katramatos, M.~Keach, A.~Parsells, O.~Saira, J.~Schiff,
  P.~Svihra, T.~Tsang, and Y.~Zhang, \enquote{Quantum-assisted optical
  interferometers: instrument requirements,} in \emph{Optical and Infrared
  Interferometry and Imaging VII,}  A.~M{\'{e}}rand, S.~Sallum, and P.~G.
  Tuthill, eds. (2020), Proc. SPIE.

\bibitem{Keach2022}
M.~Keach, S.~Bellavia, Z.~Chen, J.~Crawford, D.~Dolzhenko, E.~Figueroa,
  A.~Mueninghoff, A.~Nomerotski, J.~Schiff, R.~Simovitch, A.~Slosar,
  P.~Stankus, and S.~Vintskevich, \enquote{Increasing baselines and precision
  of optical interferometers using two-photon interference effects,} in
  \emph{Optical and Infrared Interferometry and Imaging {VIII},}
  A.~M{\'{e}}rand, S.~Sallum, and J.~Sanchez-Bermudez, eds. ({SPIE}, 2022).

\bibitem{HBT_original}
R.~H. Brown and R.~Twiss, \enquote{A test of a new type of stellar
  interferometer on sirius,} {\protect\JournalTitle{Nature}} \textbf{178},
  1046--1048 (1956).

\bibitem{HOM_effect}
C.~K. Hong, Z.~Y. Ou, and L.~Mandel, \enquote{Measurement of subpicosecond time
  intervals between two photons by interference,} {\protect\JournalTitle{Phys.
  Rev. Lett.}} \textbf{59}, 2044--2046 (1987).

\bibitem{NohMandel1991}
J.~W. Noh, A.~Foug\`eres, and L.~Mandel, \enquote{Measurement of the quantum
  phase by photon counting,} {\protect\JournalTitle{Phys. Rev. Lett.}}
  \textbf{67}, 1426--1429 (1991).

\bibitem{NohMandel1992a}
J.~W. Noh, A.~Foug\`eres, and L.~Mandel, \enquote{Operational approach to the
  phase of a quantum field,} {\protect\JournalTitle{Phys. Rev. A}} \textbf{45},
  424--442 (1992).

\bibitem{NohMandel1992b}
J.~W. Noh, A.~Foug\`eres, and L.~Mandel, \enquote{Further investigations of the
  operationally defined quantum phase,} {\protect\JournalTitle{Phys. Rev. A}}
  \textbf{46}, 2840--2852 (1992).

\bibitem{NohMandel1993}
J.~W. Noh, A.~Foug\`eres, and L.~Mandel, \enquote{Measurements of the
  probability distribution of the operationally defined quantum phase
  difference,} {\protect\JournalTitle{Phys. Rev. Lett.}} \textbf{71},
  2579--2582 (1993).

\bibitem{Sensors2020_Nomerotski}
A.~Nomerotski, M.~Keach, P.~Stankus, P.~Svihra, and S.~Vintskevich,
  \enquote{Counting of hong-ou-mandel bunched optical photons using a fast
  pixel camera,} {\protect\JournalTitle{Sensors}} \textbf{20} (2020).

\bibitem{PRA_Jordan2022}
K.~M. Jordan, R.~A. Abrahao, and J.~S. Lundeen, \enquote{Quantum metrology
  timing limits of the hong-ou-mandel interferometer and of general two-photon
  measurements,} {\protect\JournalTitle{Phys. Rev. A}} \textbf{106}, 063715
  (2022).

\bibitem{Bouchard_HOM_review}
F.~Bouchard, A.~Sit, Y.~Zhang, R.~Fickler, F.~M. Miatto, Y.~Yao, F.~Sciarrino,
  and E.~Karimi, \enquote{Two-photon interference: the hong–ou–mandel
  effect,} {\protect\JournalTitle{Reports on Progress in Physics}} \textbf{84},
  012402 (2020).

\bibitem{harvard1}
E.~T. Khabiboulline, J.~Borregaard, K.~De~Greve, and M.~D. Lukin,
  \enquote{Quantum-assisted telescope arrays,} {\protect\JournalTitle{Phys.
  Rev. A}} \textbf{100}, 022316 (2019).

\bibitem{harvard2}
E.~T. Khabiboulline, J.~Borregaard, K.~De~Greve, and M.~D. Lukin,
  \enquote{Optical interferometry with quantum networks,}
  {\protect\JournalTitle{Phys. Rev. Lett.}} \textbf{123}, 070504 (2019).

\bibitem{Brown2021}
M.~Brown, V.~Thiel, M.~Allgaier, M.~Raymer, B.~Smith, P.~Kwiat, and J.~Monnier,
  \enquote{Interferometry-based astronomical imaging using nonlocal
  interference with single-photon states,} in \emph{Frontiers in Optics $+$
  Laser Science 2021,}  ({OSA}, 2021).

\bibitem{Brown2022}
M.~Brown, V.~Thiel, M.~Allgaier, M.~Raymer, B.~Smith, P.~Kwiat, and J.~Monnier,
  \enquote{Long-baseline interferometry using single photon states as a
  non-local oscillator,} in \emph{Quantum Computing, Communication, and
  Simulation {II},}  P.~R. Hemmer and A.~L. Migdall, eds. ({SPIE}, 2022).

\bibitem{FIO2021Kwiat}
D.~Diaz, Y.~Zhang, V.~O. Lorenz, and P.~G. Kwiat, \enquote{Emulating
  quantum-enhanced long-baseline interferometric telescopy,} in \emph{Frontiers
  in Optics $+$ Laser Science 2021,}  (Optica Publishing Group, 2021), p.
  FTh6D.7.

\bibitem{OSAOptica2_Kwiat_2022}
M.~Brown, V.~Thiel, M.~Allgaier, M.~Raymer, B.~Smith, P.~Kwiat, and J.~Monnier,
  \enquote{Proof-of-principle laboratory demonstration of long-baseline
  interferometric imaging using distributed single-photons,} in \emph{Quantum
  2.0 Conference and Exhibition,}  (Optica Publishing Group, 2022), p. QM3C.1.

\bibitem{SPDC_general}
A.~Christ, A.~Fedrizzi, H.~H{\"u}bel, T.~Jennewein, and C.~Silberhorn,
  \enquote{Parametric down-conversion,} in \emph{Experimental Methods in the
  Physical Sciences,}  vol.~45 (Elsevier, 2013), pp. 351--410.

\bibitem{Gasparini17}
L.~Gasparini, B.~Bessire, M.~Unternahrer, A.~Stefanov, D.~Boiko, M.~Perenzoni,
  and D.~Stoppa, \enquote{{SUPERTWIN}: towards 100kpixel {CMOS} quantum image
  sensors for quantum optics applications,} in \emph{Quantum Sensing and Nano
  Electronics and Photonics {XIV},}  M.~Razeghi, ed. (2017), Proc. SPIE.

\bibitem{Perenzoni16}
M.~Perenzoni, L.~Pancheri, and D.~Stoppa, \enquote{Compact spad-based pixel
  architectures for time-resolved image sensors,}
  {\protect\JournalTitle{Sensors}} \textbf{16} (2016).

\bibitem{Lee18}
M.~Lee and E.~Charbon, \enquote{Progress in single-photon avalanche diode image
  sensors in standard {CMOS}: From two-dimensional monolithic to
  three-dimensional-stacked technology,} {\protect\JournalTitle{Japanese
  Journal of Applied Physics}} \textbf{57}, 1002A3 (2018).

\bibitem{Divochiy08}
A.~Divochiy, F.~Marsili, D.~Bitauld, A.~Gaggero, R.~Leoni, F.~Mattioli,
  A.~Korneev, V.~Seleznev, N.~Kaurova, O.~Minaeva,
  G.~Gol{\textquotesingle}tsman, K.~G. Lagoudakis, M.~Benkhaoul, F.~L{\'{e}}vy,
  and A.~Fiore, \enquote{Superconducting nanowire photon-number-resolving
  detector at telecommunication wavelengths,} {\protect\JournalTitle{Nature
  Photonics}} \textbf{2}, 302--306 (2008).

\bibitem{Zhu20}
D.~Zhu, M.~Colangelo, C.~Chen, B.~A. Korzh, F.~N.~C. Wong, M.~D. Shaw, and
  K.~K. Berggren, \enquote{Resolving photon numbers using a superconducting
  nanowire with impedance-matching taper,} {\protect\JournalTitle{Nano
  Letters}}  (2020).

\bibitem{Korzh20}
B.~Korzh, Q.~Zhao, J.~P. Allmaras, S.~Frasca, T.~M. Autry, E.~A. Bersin, A.~D.
  Beyer, R.~M. Briggs, B.~Bumble, M.~Colangelo, G.~M. Crouch, A.~E. Dane,
  T.~Gerrits, A.~E. Lita, F.~Marsili, G.~Moody, C.~Pe{\~{n}}a, E.~Ramirez,
  J.~D. Rezac, N.~Sinclair, M.~J. Stevens, A.~E. Velasco, V.~B. Verma, E.~E.
  Wollman, S.~Xie, D.~Zhu, P.~D. Hale, M.~Spiropulu, K.~L. Silverman, R.~P.
  Mirin, S.~W. Nam, A.~G. Kozorezov, M.~D. Shaw, and K.~K. Berggren,
  \enquote{Demonstration of sub-3 ps temporal resolution with a superconducting
  nanowire single-photon detector,} {\protect\JournalTitle{Nature Photonics}}
  \textbf{14}, 250--255 (2020).

\bibitem{lmfit}
M.~Newville, R.~Otten, A.~Nelson, A.~Ingargiola, T.~Stensitzki, D.~Allan,
  A.~Fox, F.~Carter, Michał, R.~Osborn, D.~Pustakhod, lneuhaus, S.~Weigand,
  Glenn, C.~Deil, Mark, A.~L.~R. Hansen, G.~Pasquevich, L.~Foks, N.~Zobrist,
  O.~Frost, A.~Beelen, Stuermer, azelcer, A.~Hannum, A.~Polloreno, J.~H.
  Nielsen, S.~Caldwell, A.~Almarza, and A.~Persaud, \enquote{lmfit/lmfit-py:
  1.0.3,} {\protect\JournalTitle{Zenodo}}  (2021).

\bibitem{Zhi2022}
Z.~Chen, A.~Nomerotski, A.~Slo{\v{z}}ar, P.~Stankus, and S.~Vintskevitch,
  \enquote{Astrometry in two-photon interferometry using earth rotation fringe
  scan,} {\protect\JournalTitle{arXiv preprint arXiv:2205.09091}}  (2022).

\bibitem{Serafini2017}
A.~Serafini, \emph{Quantum continuous variables: a primer of theoretical
  methods} (CRC press, 2017).

\bibitem{Xia21}
Y.~Xia, W.~Li, W.~Clark, D.~Hart, Q.~Zhuang, and Z.~Zhang, \enquote{Entangled
  sensor networks empowered by machine learning,} in \emph{Optical Fiber
  Communication Conference (OFC) 2021,}  (Optica Publishing Group, 2021), p.
  Th3A.4.

\bibitem{VintskevichPRA2019}
S.~V. Vintskevich, D.~A. Grigoriev, and S.~N. Filippov, \enquote{Effect of an
  incoherent pump on two-mode entanglement in optical parametric generation,}
  {\protect\JournalTitle{Phys. Rev. A}} \textbf{100}, 053811 (2019).

\bibitem{Brady2022}
A.~J. Brady, C.~Gao, R.~Harnik, Z.~Liu, Z.~Zhang, and Q.~Zhuang,
  \enquote{Entangled sensor-networks for dark-matter searches,}
  {\protect\JournalTitle{PRX Quantum}} \textbf{3}, 030333 (2022).

\bibitem{wu2022entanglement}
B.-H. Wu, S.~Guha, and Q.~Zhuang, \enquote{Entanglement-assisted multi-aperture
  pulse-compression radar for angle resolving detection,}
  {\protect\JournalTitle{arXiv preprint arXiv:2207.10881}}  (2022).

\bibitem{Zhang2019}
Q.~Zhuang and Z.~Zhang, \enquote{Physical-layer supervised learning assisted by
  an entangled sensor network,} {\protect\JournalTitle{Phys. Rev. X}}
  \textbf{9}, 041023 (2019).

\bibitem{cox2022transceiver}
A.~Cox, Q.~Zhuang, C.~Gagatsos, B.~Bash, and S.~Guha, \enquote{Transceiver
  designs to attain the entanglement assisted communications capacity,}
  {\protect\JournalTitle{arXiv preprint arXiv:2208.07979}}  (2022).

\bibitem{Krenn_review}
M.~Krenn, J.~Landgraf, T.~Foesel, and F.~Marquardt, \enquote{Artificial
  intelligence and machine learning for quantum technologies,}
  {\protect\JournalTitle{arXiv:2208.03836}}  (2022).

\bibitem{vintskevich2022}
S.~Vintskevich, N.~Bao, A.~Nomerotski, P.~Stankus, and D.~Grigoriev,
  \enquote{Classification of four-qubit entangled states via machine learning,}
  {\protect\JournalTitle{arXiv preprint arXiv:2205.11512}}  (2022).

\end{thebibliography}


\begin{thebibliography}{10}
\newcommand{\enquote}[1]{``#1''}

\bibitem{Gasparini17}
L.~Gasparini, B.~Bessire, M.~Unternahrer, A.~Stefanov, D.~Boiko, M.~Perenzoni,
  and D.~Stoppa, \enquote{{SUPERTWIN}: towards 100kpixel {CMOS} quantum image
  sensors for quantum optics applications,} in \emph{Quantum Sensing and Nano
  Electronics and Photonics {XIV},}  M.~Razeghi, ed. (2017), Proc. SPIE.

\bibitem{Perenzoni16}
M.~Perenzoni, L.~Pancheri, and D.~Stoppa, \enquote{Compact spad-based pixel
  architectures for time-resolved image sensors,}
  {\protect\JournalTitle{Sensors}} \textbf{16} (2016).

\bibitem{Lee18}
M.~Lee and E.~Charbon, \enquote{Progress in single-photon avalanche diode image
  sensors in standard {CMOS}: From two-dimensional monolithic to
  three-dimensional-stacked technology,} {\protect\JournalTitle{Japanese
  Journal of Applied Physics}} \textbf{57}, 1002A3 (2018).

\bibitem{Morimoto20}
K.~Morimoto, A.~Ardelean, M.~Wu, A.~C. Ulku, I.~M. Antolovic, C.~Bruschini, and
  E.~Charbon, \enquote{Megapixel time-gated {SPAD} image sensor for 2d and 3d
  imaging applications,} {\protect\JournalTitle{Optica}} \textbf{7}, 346
  (2020).

\bibitem{Divochiy08}
A.~Divochiy, F.~Marsili, D.~Bitauld, A.~Gaggero, R.~Leoni, F.~Mattioli,
  A.~Korneev, V.~Seleznev, N.~Kaurova, O.~Minaeva,
  G.~Gol{\textquotesingle}tsman, K.~G. Lagoudakis, M.~Benkhaoul, F.~L{\'{e}}vy,
  and A.~Fiore, \enquote{Superconducting nanowire photon-number-resolving
  detector at telecommunication wavelengths,} {\protect\JournalTitle{Nature
  Photonics}} \textbf{2}, 302--306 (2008).

\bibitem{Zhu20}
D.~Zhu, M.~Colangelo, C.~Chen, B.~A. Korzh, F.~N.~C. Wong, M.~D. Shaw, and
  K.~K. Berggren, \enquote{Resolving photon numbers using a superconducting
  nanowire with impedance-matching taper,} {\protect\JournalTitle{Nano
  Letters}}  (2020).

\bibitem{Korzh20}
B.~Korzh, Q.~Zhao, J.~P. Allmaras, S.~Frasca, T.~M. Autry, E.~A. Bersin, A.~D.
  Beyer, R.~M. Briggs, B.~Bumble, M.~Colangelo, G.~M. Crouch, A.~E. Dane,
  T.~Gerrits, A.~E. Lita, F.~Marsili, G.~Moody, C.~Pe{\~{n}}a, E.~Ramirez,
  J.~D. Rezac, N.~Sinclair, M.~J. Stevens, A.~E. Velasco, V.~B. Verma, E.~E.
  Wollman, S.~Xie, D.~Zhu, P.~D. Hale, M.~Spiropulu, K.~L. Silverman, R.~P.
  Mirin, S.~W. Nam, A.~G. Kozorezov, M.~D. Shaw, and K.~K. Berggren,
  \enquote{Demonstration of sub-3 ps temporal resolution with a superconducting
  nanowire single-photon detector,} {\protect\JournalTitle{Nature Photonics}}
  \textbf{14}, 250--255 (2020).

\bibitem{Andrei_camera2022}
A.~Nomerotski, M.~Chekhlov, D.~Dolzhenko, R.~Glazenborg, B.~Farella, M.~Keach,
  R.~Mahon, D.~Orlov, and P.~Svihra, \enquote{Intensified tpx3cam, a fast
  data-driven optical camera with nanosecond timing resolution for single
  photon detection in quantum applications,}  (2022).

\bibitem{Nomerotski20}
A.~Nomerotski, P.~Stankus, A.~Slo{\v{z}}ar, S.~Vintskevich, S.~Andrewski, G.~A.
  Carini, D.~Dolzhenko, D.~England, E.~V. Figueroa, S.~Gera, J.~Haupt,
  S.~Herrmann, D.~Katramatos, M.~Keach, A.~Parsells, O.~Saira, J.~Schiff,
  P.~Svihra, T.~Tsang, and Y.~Zhang, \enquote{Quantum-assisted optical
  interferometers: instrument requirements,} in \emph{Optical and Infrared
  Interferometry and Imaging VII,}  A.~M{\'{e}}rand, S.~Sallum, and P.~G.
  Tuthill, eds. (2020), Proc. SPIE.

\bibitem{GLPrepr}
K.~E. Cahill and R.~J. Glauber, \enquote{Density operators and quasiprobability
  distributions,} {\protect\JournalTitle{Phys. Rev.}} \textbf{177}, 1882--1902
  (1969).

\bibitem{mandel1995optical}
L.~Mandel and E.~Wolf, \emph{Optical coherence and quantum optics} (Cambridge
  university press, 1995).

\bibitem{feynman2010quantum}
R.~P. Feynman, A.~R. Hibbs, and D.~F. Styer, \emph{Quantum mechanics and path
  integrals} (Courier Corporation, 2010).

\bibitem{glauber2007quantum}
R.~J. Glauber, \emph{Quantum theory of optical coherence: selected papers and
  lectures} (John Wiley \& Sons, 2007).

\end{thebibliography}
\bibstyle{opticajnl} 

\end{document}


\maketitle

\section{Experimental Setup and Methods}
\label{sec:setup}

The experimental setup utilized in the measurements is shown in the right part of Figure 1 in the main text. Here we provide a detailed description of the setup, which includes two argon lamps, standard optical elements, single-photon detectors, and time digital converter (TDC).

Newport 6030 Argon spectral calibration lamps operated with 10~mA DC current, acting as quasi-thermal sources, were used as simulated starlight. Photons from these lamps passed through narrow band filters with a center wavelength of 794.9~nm and 1~nm FWHM to isolate the 794.82~nm spectral line. Then the photons were coupled into single-mode fibers and directed to linear polarizers and to 50:50 non-polarizing beamsplitters. 

The reflected light from the beamsplitters was then passed through motor-controlled phase shifters, made of glass wedges, which can be moved laterally to the beam direction. The wedge angle is $1^\circ$ and the glass refractive index around the wavelength of interest is 1.51, which translates to the phase advance of about 9~nm per one micron motor step. Following that, the beam was directed into another beamsplitter, for mixing with photons from the opposite argon lamp, arriving from another beamsplitter through the transmitted path. The light from beamsplitter's output was then coupled into single-mode fibers and read out by detectors with single-photon sensitivity. 

Two types of detectors were used, single-photon avalanche detectors (SPAD) and superconducting nanowire single-photon detectors (SNSPD).
SPADs are silicon-based devices with internal amplification in the diode junction. They produce fast pulses operating with single photon sensitivity \cite{Gasparini17, Perenzoni16, Lee18}.  SPAD devices are capable of achieving timing resolution as good as 10~ps for single channel devices.  These devices also have good scalability, with multichannel imagers already reported \cite{Gasparini17, Morimoto20}. We used the Thorlabs SPDMA devices with photon detection efficiency of 47\% for 795~nm and timing resolution of 200~ps (rms).

SNSPD is a technology with excellent quantum efficiency and time resolution. We utilized a commercial unit (SingleQuantum EOS), which provides four infrared-sensitive channels made of serpentine wires kept at superconducting temperatures by a helium compressor. Voltage signals are induced in the detection circuit when a photon deposits its energy in the vicinity of the wire and momentarily breaks the superconductivity \cite{Divochiy08, Zhu20, Korzh20}. The SNSPDs used for the measurements here had timing resolution of 100~ps (rms). 
The SPAD/SNSPD digital output signals are then read out by a quTAG TDC module, which timestamps individual photons with 7~ps timing resolution.

The interferometer was aligned using a 780~nm continuous-wave (CW) laser with resulting optical throughput (up to before the detectors) for the input flux of about 50\%. Special care was taken to equalize the rates from two lamps and between the four output channels. Typical single photon rates for the argon lamps operating at full power with SNSPD detectors were about 200-400~k counts per second per channel depending on the configuration.

To verify the phase stability we reconfigured the setup to a single photon interferometer by sending one of the argon lamp beams to a 50:50 beamsplitter and feeding the two resulting beams to the two inputs of the interferometer. In this case, the output rates directly depend on the phase difference between the two interferometer arms since it behaves as a classical Mach-Zehnder interferometer. This allowed us to determine that the phase stability was very good, of the order of 10 hours so much longer than the duration of our experiments.

We also describe here an algorithm, which was deployed to remove the afterpulses to avoid the double-counting of single photon hits. Afterpulses is a detector effect that results in the appearance of pulses only after a short time following the primary pulse \cite{Andrei_camera2022}. For our measurements, we saw the afterpulses only for the SNSPD signals. It is likely that their origin was in the SNSPD frontend electronics rather than in the nanowire sensor itself. We studied the SNSPD afterpulsing in detail in our previous work \cite{Nomerotski20}. During the initial analysis steps the afterpulses were removed by ignoring any hits appearing closer than 30~ns to the original pulse \cite{Nomerotski20}. Similar studies were performed for SPAD detectors with the conclusion that the SPAD digital signals at the output of the device did not have afterpulses.

\section{Theoretical considerations and derivations}\label{sec:Remarks}

The theoretical description for the interferometer setup in Figure 1 in the main text can be derived employing the Glauber – Sudarshan P-representation \cite{GLPrepr,mandel1995optical}, which provides a natural description of transformations in the beamsplitters and phase shifters of a joint state produced by thermal sources. We believe that the quantum description of our observations is very instructive as it provides a clear link between the mode indistinguishability and quantum interference phenomena through alternative paths in quantum mechanics \cite{feynman2010quantum,glauber2007quantum}. 

In accordance with the output channel numbering in the right part of Figure 1 in the main text, the joint quantum state of light is described as follows \cite{mandel1995optical}:
\begin{equation}\label{joint_state}
\hat{\varrho}_{in} = {\ket{0}\bra{0}}_{1,in}\otimes{[\hat{\varrho}_{2,H}\otimes\hat{\varrho}_{2,V}]}_{in}\otimes{\ket{0}\bra{0}}_{3,in}\otimes{[\hat{\varrho}_{4,H}\otimes\hat{\varrho}_{4,V}]}_{in},
\end{equation}
where the indices $H,V$ correspond to the horizontal and vertical polarization degrees of freedom for the resulting {\it{thermal states}} of spatial modes characterized by indices $2,4$ after transformation via polarizers. Note, to avoid cumbersome expressions we labeled all input modes by the label $in$. ${\ket{0}\bra{0}}_{1},{\ket{0}\bra{0}}_{3}$ are the vacuum state of spatial modes $1,3$. 
It is assumed that the average photon occupation number in a given spatial-frequency mode $m,\omega$ after polarization filter is equal to:
\begin{equation}
\braket{\hat{n}_{m}(\omega)} = \braket{\hat{a}_{m,H}^{\dagger}(\omega)\hat{a}_{m,H}(\omega)} + \braket{\hat{a}_{m,V}^{\dagger}(\omega)\hat{a}_{m,V}(\omega)} = n_{m,H}(\omega)\cos^{2}(\theta_{m}) + n_{m,V}(\omega)\sin^{2}(\theta_{m}), 
\end{equation}
where $n_{m,H}(\omega)$ and $n_{m,V}(\omega)$ are the average photon occupation numbers in horizontal ($H$) and vertical ($V$) polarization modes respectively and a given spatial mode $m$
\cite{glauber2007quantum}.
The P - representation yields the form of a multimode joint initial state in (\ref{joint_state}) as: 
\begin{eqnarray}\label{eq:init_state}
&&\hat{\varrho}_{in} =
\int d^{2}{\bm{\alpha}}_{2}d^{2}{\bm{\alpha}}_{4} P\left(\bm{\alpha}_{2H},\bm{\alpha}_{2V}\right)P\left(\bm{\alpha}_{4H},\bm{\alpha}_{4V}\right)\nonumber\\
&&{\ket{0}\bra{0}}_{1, in}\otimes{[\ket{\bm{\alpha}_{2H},\bm{\alpha}_{2V}}\bra{\bm{\alpha}_{2H},\bm{\alpha}_{2V}}]}_{in}\otimes{\ket{0}\bra{0}}_{3,in}\otimes{[\ket{\bm{\alpha}_{4H},\bm{\alpha}_{4V}}\bra{\bm{\alpha}_{4H},\bm{\alpha}_{4V}}]}_{in},
\end{eqnarray}
where the state $\ket{\bm{\alpha}}$ represents a short notation for the multimode coherent state (for polarization and frequency modes) such that $\hat{a}_{mp}(\omega)\ket{\bm{\alpha}} = \alpha_{mp}(\omega)\ket{\bm{\alpha}}, mp \in \{mH,mV\}, \ m \in \{1,2,3,4\} $ in the above.
We denoted a bosonic annihilation operator of particular mode as $\hat{a}_{pm}(\omega)$. The corresponding P-function of the thermal state with spatial modes labelled by indices $m = 2~\&~4$ have the following form:
\begin{equation}\label{eq:P-representation}
    P\left(\bm{\alpha}_{mH},\bm{\alpha}_{mV}\right) = \prod_{\omega,\omega^{'}}\frac{1}{\pi^{2}{n_{mH}(\omega)}{n_{mV}(\omega^{'})}}e^{-\frac{|\alpha_{mH}(\omega)|^{2}}{{n_{mH}(\omega)}}}e^{-\frac{|\alpha_{mV}(\omega^{'})|^{2}}{{n_{mV}(\omega^{'})}}},  \ m \in\{2,4\}.
\end{equation}
The $P$ - representation is a useful tool for the theoretical description, especially in the case of thermal statistics of quantum states owing to a significant simplification of expressions and direct correspondence to the classical optics. Note that there is a one-to-one correspondence between the input modes and operators describing the quantum states and observables related directly to the selected modes. 
The initial state of light is transformed via action of the unitary quantum channel describing the action of nonpolarizing beamsplitters ($\hat{U}$) and phase shifters ($\hat{V}$):
\begin{equation}\label{eq:trnasformations}
\hat{\varrho}_{out} = (\hat{U}_{12}\otimes\hat{U}_{34})\hat{V}_{3}\hat{V}_{1}(\hat{U}_{14}\otimes\hat{U}_{23})\hat{\varrho}_{in}[(\hat{U}_{12}\otimes\hat{U}_{34})\hat{V}_{3}\hat{V}_{1}(\hat{U}_{14}\otimes\hat{U}_{23})]^{\dagger}.
\end{equation}
The action of this unitary channel can be specified by the transformation: 
\begin{equation}
\label{eq:transformations_ext}
\begin{bmatrix} 
     \hat{a}_{m}\\
     \hat{a}_{k}
\end{bmatrix} = \hat{U}_{mk}\begin{bmatrix} 
     \hat{a}^{in}_{m}\\
     \hat{a}^{in}_{k}
\end{bmatrix} = \begin{bmatrix} 
     \frac{1}{\sqrt{2}} & \frac{1}{\sqrt{2}}\\
     -\frac{1}{\sqrt{2}} & \frac{1}{\sqrt{2}}
\end{bmatrix}\begin{bmatrix} 
     \hat{a}^{in}_{m}\\
     \hat{a}^{in}_{k},
\end{bmatrix}
\end{equation}
where $m$ and $k$ are the corresponding spatial modes. The following equalities hold:
\begin{eqnarray}\label{eq:transform_trick}
&&\hat{V}(\delta)_{m}\hat{U}_{mk}\ket{\bm{\alpha}_{mp}}_{m,in}\ket{\bm{\alpha}_{kp}}_{k,in} = \ket{e^{i\delta}\frac{\bm{\alpha}_{mp} + \bm{\alpha}_{kp}}{\sqrt{2}}}_{m}\ket{\frac{-\bm{\alpha}_{mp} + \bm{\alpha}_{kp}}{\sqrt{2}}}_{k}, p \in \{H,V\} \\ &&\hat{V}(\delta)_{m}\hat{U}_{mk}\ket{\bm{\alpha}_{mH(V)}}_{m,in}\ket{\bm{\alpha}_{kV(H)}}_{k,in} =\nonumber\\
&& = \ket{e^{i\delta}\frac{\bm{\alpha}_{mH(V)}}{\sqrt{2}}}_{m}\ket{e^{i\delta}\frac{\bm{\alpha}_{kV(H)}}{\sqrt{2}}}_{m}\ket{-\frac{\bm{\alpha}_{mH(V)}}{\sqrt{2}}}_{k}\ket{\frac{\bm{\alpha}_{kV(H)}}{\sqrt{2}}}_{k}
\end{eqnarray}

Thus, taking into account \eqref{eq:init_state} - \eqref{eq:transform_trick} the resulting state (\ref{eq:trnasformations}) can be rewritten as:
\begin{eqnarray}
\label{eq:result}
&&\hat{\varrho}_{out} = \int d^{2}{\bm{\alpha}}_{2}d^{2}{\bm{\alpha}}_{4} P\left(\bm{\alpha}_{2}\right)P\left(\bm{\alpha}_{4}\right)\nonumber\\
&&\left({\ket{\frac{\bm{\alpha}_{4} + e^{i\delta_{1}}\bm{\alpha}_{2}}{2}}}_{1 \ 1}{\bra{\frac{\bm{\alpha}_{4} + e^{i\delta_{1}}\bm{\alpha}_{2}}{2}}}\right)\otimes\left({\ket{\frac{\bm{\alpha}_{4} + e^{i\delta_{2}}\bm{\alpha}_{2}}{2}}}_{2 \ 2}{\bra{\frac{\bm{\alpha}_{4} + e^{i\delta_{2}}\bm{\alpha}_{2}}{2}}}\right)\otimes \nonumber \\
&&\left({\ket{\frac{\bm{\alpha}_{4} + e^{i\delta_{3}}\bm{\alpha}_{2}}{2}}}_{3 \ 3}{\bra{\frac{\bm{\alpha}_{4} + e^{i\delta_{3}}\bm{\alpha}_{2}}{2}}}\right)\otimes\left({\ket{\frac{\bm{\alpha}_{4} + e^{i\delta_{4}}\bm{\alpha}_{2}}{2}}}_{4 \ 4}{\bra{\frac{\bm{\alpha}_{4} + e^{i\delta_{4}}\bm{\alpha}_{2}}{2}}}\right),
\end{eqnarray}
where $\delta_{1,3} = (\Delta_{2,3} -\Delta_{1,4})\omega/c; ~\delta_{2,4} = (\Delta_{2,3} -\Delta_{1,4})\omega/c + \pi$, $\Delta_{j}$ - a phase shift along direction of a given mode $j \in \{1,2,3,4\}$. For the sake of notation simplicity, we absorbed the polarisation degrees of freedom assuming $\bm{\ket{\alpha}} \equiv \bm{\ket{\alpha}}_{H}\bm{\ket{\alpha}}_{V}$. 

In our experiments, we measure the coincidence rates between different pairs of detectors that register light in different spatial modes. Theoretically, such measurements are described by the intensity (second-order) correlation function $\Gamma_{jp,kq}(\Delta t)$ with spatial modes labeled with $j\neq k\in \{1,2,3,4\}$ and polarization modes labeled with $ \{p,q\} \in \{H,V\}$. For the degree of second-order coherence the corresponding operator $\hat{O}_{jp,kq}(\Delta t)$ describing the intensity correlations reads:
\begin{equation}\label{corr_func}
\Gamma_{jp,kq}(\Delta t) = {\rm{tr}}\left(\hat{O}(\Delta t)_{jk}\hat{\varrho}_{out}\right) = {\rm{tr}}\left(\hat{\vec{E}}^{\dagger}_{jp}(t)\hat{\vec{E}}^{\dagger}_{kq}(t+\Delta t)\hat{\vec{E}}_{kq}(t+\Delta t)\hat{\vec{E}}_{jp}(t)\hat{\varrho}_{out}\right),
\end{equation}
where ${\rm{tr}}$ is the full trace operation. The normalized correlation function $g_{jp,kq}^{(2)}(\Delta t)$ (also commonly named as $g^{(2)}$-factor or normalized correlation function) can be expressed as follows:
\begin{equation}\label{corr_int}
g^{(2)}_{jp,kq}(\Delta t) = \frac{\Gamma_{jp,kq}(\Delta t)}{{\rm{tr}}\left(\hat{\varrho}_{out}\hat{\vec{E}}^{\dagger}_{jp}(t)\hat{\vec{E}}_{jp}(t)\right){\rm{tr}}\left(\hat{\varrho}_{out}\hat{\vec{E}}^{\dagger}_{kq}(t + \Delta t)\hat{\vec{E}}_{kq}(t + \Delta t)\right)} 
\end{equation}
The detectors are labeled in accordance with corresponding spatial modes. One can use a standard one-dimensional expansion of field operators in the time-frequency domain:
\begin{equation}
\hat{\vec{E}}_{m}(t) = \int \hat{a}_{Hm}(\omega)\vec{e}_{H}e^{-i\omega t}d\omega + \int \hat{a}_{Vm}(\omega)\vec{e}_{V}e^{-i\omega t}d\omega 
\end{equation}

Note that if one assumes that the polarization degree of freedom is not changing and the same is true for all spatial modes, one can omit indices $p,q$ for simplicity.
To analyze specific measurements results one can utilize several simplifications in  \eqref{corr_func} and \eqref{corr_int}. In Section 3.1 of the main text we analyzed the HBT peaks for setups with two different configurations of polarization modes $VV$ and $VH$. For this cases \eqref{corr_int} yields:
\begin{eqnarray}\label{HBT_HV}
&& \Gamma_{jV,kV}(\Delta t,t) = 2n_{4V}^{2}f_{4V}(jk,\Delta t) + 2n_{4V}^{2}f_{2V}(jk,\Delta t) + \nonumber\\
&& + 2n_{4V}n_{2V}\big[1 + \tilde{f}_{4V}(jk,\Delta t)\tilde{f}_{2V}(jk,\Delta t)\cos{(\delta_{ij}(t))}\big] \\
&&\Gamma_{jV,kH}(\Delta t,t) =  2n_{4V}^{2}f_{4V}(\Delta t) + 2n_{2H}^{2}f_{2H}(\Delta t) + 2n_{4V}n_{2H},
\end{eqnarray}
where the phase factor $\delta_{ij}(t) = (\delta_{i} - \delta_{j})  - \frac{2\pi t}{T}$ has contributions from constant phase shifts $(\delta_{i} - \delta_{j})$ and a slowly varying phase shift with period $T$ induced with the phase shifters, see Fig.1 in the main text. As usual $n_{mp}$ is the average photon number in input modes. Functions $f_{mp}(jk,\Delta t),m\in\{2,4\},p\in\{H,V\}$ are defined by the source spectrum and overall phase shifts between the input sources ($2,4$) and pairs of detectors, labeled by $i,j$. We use the Gaussian spectrum to model such functions: $f_{mp}(jk,\Delta t) = C_{mp}e^{-\frac{(t_{jk} - \Delta t)^{2}}{2\sigma^{2}}}$.  In Sec.3.2 of the main text we analyze the oscillation of coincidence rates by estimating $\Gamma_{jV,kV}(\Delta t,t)$ and $\Gamma_{jV,kH}(\Delta t,t)$ within a specific coincidence time window near the peak maximum and picking a corresponding time bin size. This is equivalent to averaging in the time domain; we can replace $\Gamma_{jV,kV(H)}(\Delta t,t)$  with averaged functions $\langle\Gamma_{jV,kV(H)}(\Delta t,t)\rangle$. In this case the functions $f_{mp}(jk,\Delta t)$ can be replaced by constants and the resulting $\langle\Gamma_{jV,kV(H)}(\Delta t,t)\rangle$ gets the following functional form $\langle\Gamma_{ip_{i}jp_{j}}(t)\rangle = \langle A \rangle_{ip_{i}jp_{j},} + \langle B\rangle_{ip_{i}jp_{j}}\cos{\left(\frac{2\pi}{T}t-\Delta\delta_{ij} \right)}$. We denoted here labels of polarization as $p_{i}$ and $p_{i}$ associated with spatial modes $i$ and $j$. 

Note that the quantum interference effects (and consequently oscillation pattern governed by $\cos{(\delta_{ij}(t))} = \cos{( (\delta_{i} - \delta_{j})  - \frac{2\pi t}{T})}$) vanish if modes produced by sources are distinguishable (e.g. $p_{i} = V$ and $p_{j} = H$), which we see clearly from Eqs.(S13) and (S14). We emphasize the importance of mode indistinguishability focusing on the polarization degree of freedom in 
the derivation below.

Let us explicitly calculate \eqref{corr_int} within the quasi-monochromatic approximation $\Delta\omega T_{res} \ll \omega_{0}$ and by assuming $\Delta\omega T_{res} \ll 1$, where $T_{res}$ is the detector temporal resolution. Substituting \eqref{eq:result} and \eqref{corr_func} in \eqref{corr_int} and taking it to the limit $\Delta \omega \longrightarrow 0 $ we obtain:
\begin{eqnarray}\label{g2_result}
&&g^{(2)}_{jk} = 2\frac{\overline{n}_{4H}^{2} + \overline{n}_{2H}^{2}+ \overline{n}_{4V}^{2} + \overline{n}_{2V}^{2} +\overline{n}_{4H}\overline{n}_{2H} + \overline{n}_{4V}\overline{n}_{2V} + (\overline{n}_{4H} + \overline{n}_{2H})(\overline{n}_{4V} + \overline{n}_{2V})}{\left(\overline{n}_{4H}+\overline{n}_{2H}+\overline{n}_{4V}+\overline{n}_{2V}\right)^{2}} + \nonumber \\
&& + 2\frac{\overline{n}_{4H}\overline{n}_{2H} + \overline{n}_{4V}\overline{n}_{2V}}{\left(\overline{n}_{4H}+\overline{n}_{2H}+\overline{n}_{4V}+\overline{n}_{2V}\right)^{2}}\cos{\left(\delta_{j} - \delta_{k}\right)}.
\end{eqnarray}
Let us also explicitly write down the interference visibility $\mathcal{V}_{theory}$:
\begin{eqnarray}
&&{\mathcal{V}_{theory}}= 
\frac{{\underset{\delta_{j},\delta_{k}}{\rm{max}}}(g_{jk}^{(2)}) - {\underset{\delta_{j},\delta_{k}}{\rm{min}}}(g_{jk}^{(2)})}{{\underset{\delta_{j},\delta_{k}}{\rm{max}}}(g_{jk}^{(2)}) + {\underset{\delta_{j},\delta_{k}}{\rm{min}}}(g_{jk}^{(2)})}= \nonumber \\
&& = \frac{\overline{n}_{4H}\overline{n}_{2H} + \overline{n}_{4V}\overline{n}_{2V}}{\overline{n}_{4H}^{2} + \overline{n}_{2H}^{2}+ \overline{n}_{4V}^{2} + \overline{n}_{2V}^{2} +\overline{n}_{4H}\overline{n}_{2H} + \overline{n}_{4V}\overline{n}_{2V} + (\overline{n}_{4H} + \overline{n}_{2H})(\overline{n}_{4V} + \overline{n}_{2V})}.
\label{eq:vis_theory_supp}
\end{eqnarray}
The limit $\Delta \omega \longrightarrow 0 $ allows us to exclude dependence on $\Delta t$ and concentrate on the polarization degree of freedom. 

Equation \eqref{eq:vis_theory_supp} is important to understand the results. The quantum interference vanishes ($\mathcal{V}_{theory} = 0$) when the polarization modes from the sources are different so the photons from the two sources are distinguishable: $\overline{n}_{4H} = 0, \overline{n}_{2H} \neq 0,\overline{n}_{4V}\neq 0, \overline{n}_{2V} = 0 $; corresponding to the right part of Figure 3 in the main text. However, the HBT effect still holds (for both modes V and H) independently for each pair of detectors $1\&2,1\&3,1\&4,2\&3,2\&4,3\&4$, see the right part of Figure 2 in the main text.

Alternatively, if  $\overline{n}_{4H} = 0, \overline{n}_{2H} = 0, \overline{n}_{4V}\neq 0, \overline{n}_{2V} \neq 0 $ the visibility $\mathcal{V}_{theory} \neq 0$ and the oscillatory behaviour is restored due to the mode indistinguishability. Note that if $\overline{n}_{4V} \approx \overline{n}_{2V}$ we observe a destructive interference for the combinations $1\&2$ and $3\&4$ due to $\cos{(\delta_{2} - \delta_{1})} = \cos{(\delta_{3} - \delta_{4})} = \cos{(\pi)} = -1$. 
Not perfect cancellation may happen due to only approximate equality $\overline{n}_{4V} \approx \overline{n}_{2V}$ and possible imperfections of photodetection.

Finally, let us analyze the visibility of oscillatory effects in coincidence detection as a function of polarization angle $\theta$, presented in Figure 6 in the main text together with theoretical fits. Below we consider two specific cases: vertical polarization of the first lamp: $\overline{n}_{4H} = 0, \overline{n}_{4V} \neq 0$ and varying polarization angle for the second lamp, measured from the initial vertical position, and the opposite case with horizontal polarization of the first lamp.  Theoretical functions can be derived from \eqref{eq:vis_theory_supp}. We can rewrite expression for visibility by using explicit expressions $\overline{n}_{2H} = n_{2H}\cos^{2}{\theta},\overline{n}_{2V} = n_{2V}\sin^{2}{\theta}$ in \eqref{eq:vis_theory_supp} and dividing it by $\overline{n}_{4V}$ in the VV case and by $\overline{n}_{4H}$ in the VH case.
\begin{eqnarray}
    && {V_{VV}(\theta)} = {\mathcal{V}_{theory}(\theta)}|_{\overline{n}_{4H} =0}^{\overline{n}_{4V} \neq 0} = \nonumber \\ 
    &&\frac{r_{VV} \sin^{2}(\theta)}{1 + r_{VV}\sin^{2}{\theta}+r_{VV}^{2}\sin^{4}{\theta}+r_{HV}^{2}\cos^{4}{\theta} + r_{HV}\cos^{2}{\theta} + r_{HV}r_{VV}\sin^{2}{\theta}\cos^{2}{\theta}} \\ 
    && {V_{VH}(\theta)} = {\mathcal{V}_{theory}(\theta)}|_{\overline{n}_{4H} \neq 0}^{\overline{n}_{4V} = 0} = \nonumber \\ 
    &&\frac{r_{HH} \cos^{2}(\theta)}{1 + r_{HH}\cos^{2}{\theta}+r_{HH}^{2}\cos^{4}{\theta}+r_{VH}^{2}\sin^{4}{\theta} + r_{VH}\sin^{2}{\theta} + r_{HH}r_{VH}\sin^{2}{\theta}\cos^{2}{\theta}},
\end{eqnarray}
where $\theta = \theta_{0} + \Delta\theta$ and $\theta_{0}$ is an initial polarizer position; $\Delta\theta$ is the polarization angle shift; we also denoted $r_{VV} = n_{2V}/n_{4V}, r_{HV} = n_{2H}/n_{4V}$ and  $r_{HH} = n_{2H}/n_{4H}, r_{VH} = n_{2V}/n_{4H}$ respectively for VV and VH configurations. 
From Figure 6 in the main text, it is clear that the theoretical fits are in good agreement with the experimental data.

\bibliography{sample}